\documentclass[twocolumn,showpacs,prb,amsmath,amssymb]{revtex4}
\usepackage{graphicx}% Include figure files
\usepackage{bm}% bold math
 
\DeclareMathOperator{\Max}{Max}
\DeclareMathOperator{\Min}{Min}
\DeclareMathOperator{\diag}{diag}
\begin{document}
%\draft 
%\tighten
%\onecolumn
%@twocolumnfalse\endcsname
\title{Qubit coherence control in a nuclear spin bath}
\author{Rogerio \surname{de Sousa}} 
\author{Neil \surname{Shenvi}}
\author{K. Birgitta \surname{Whaley}}
\affiliation{Department of
Chemistry and Pitzer Center for Theoretical Chemistry,\\ University of
California, Berkeley, CA 94720-1460}
\date{\today}
\begin{abstract}
Coherent dynamics of localized spins in semiconductors is limited by
spectral diffusion arising from dipolar fluctuation of lattice nuclear
spins.  Here we extend the semiclassical theory of spectral diffusion
for nuclear spins $I=1/2$ to the high nuclear spins relevant to the
III-V materials and show that applying successive qubit
$\pi$-rotations at a rate approximately proportional to the nuclear
spin quantum number squared ($I^2$) provides an efficient method for
coherence enhancement.  Hence robust coherent manipulation in the
large spin environments characteristic of the III-V compounds is
possible without resorting to nuclear spin polarization, provided that
the $\pi$-pulses can be generated at intervals scaling as $I^{-2}$.
\end{abstract}
\maketitle
%\vskip2pc]
%\narrowtext

\section{Introduction}

The ability to enhance coherence times of localized electron or
nuclear spin qubits in a semiconductor environment is central for the
development of a spin based solid state quantum computer
\cite{see_Awschalom_book}.  Qubit coherence in these devices appears
to be limited by the interaction with nuclear spins composing the
material \cite{khaetskii02,desousa03a}, with other mechanisms being
less important. A detailed comparison of the many mechanisms
contributing to localized spin decoherence in semiconductors is given
in Refs.~\onlinecite{desousa03a,desousa03b}. Experimental evidence of
the relative contributions of different decoherence channels is given
for silicon in Refs.~\onlinecite{abe04,tyryshkin03}.  Here we consider
qubit coherence under the strong effective magnetic fields and low
temperatures required for single spin read-out ($B>5$~Tesla and $T\sim
100$~mK, see Ref.~\onlinecite{elzerman04}).  Under these conditions,
environmental nuclear spin fluctuations result primarily from
inter-nuclear dipolar coupling, which becomes a source for
time-dependent noise in the qubit Zeeman energy and for its consequent
decoherence.  This channel for phase relaxation is usually denoted
spectral diffusion \cite{herzog56,klauder62,abe04,desousa03b}.

It was pointed out recently that substantial nuclear polarization can
suppress spectral diffusion \cite{khaetskii02,desousa03b}. However, the
requirement for nuclear polarization adds significant device overhead,
since polarizations on the order of 95\% or more are needed in order
to achieve significant coherence enhancement (See Fig.~3 of
Ref.~\onlinecite{dassarma05}). Recently, several schemes for nuclear
polarization were proposed \cite{imamoglu03}, but it is still not
clear whether these will be effective enough to compete with the
nuclear spin diffusion rate [$1-10$~KHz, see Eq.~(\ref{tnm}) below].

Several pulse sequences have been designed to average out the effect
of nuclear-nuclear dipolar coupling, most notably WAHUHA and
Lee-Goldburg schemes \cite{lee65}. Because these sequences are based
on nuclear spin excitation, they inevitably generate a drastic
time-dependent drift in the spin qubit Zeeman frequency.  The
amplitude of this drift can be as much as $50$~G (typical hyperfine
line-width for a localized electron in a III-V compound), fluctuating
at the nuclear spin rotation frequency (typically $10^4-10^6$~Hz
depending on the rf field intensity).  The effect of these pulses is
additional field fluctuation which adds to the decoherence rate and
makes qubit control impossible. Under such a protocol, the fidelity of
quantum operations would be drastically affected in a negative sense.
Conversely, the nuclear spins causing spectral diffusion are subject
to an inhomogeneous field arising from the interaction with the qubit.
This leads to a wide spectrum of nuclear resonance frequencies,
requiring broad-band excitation in order to perform nuclear spin
rotation. Broad-band excitation usually requires high power
deposition, which is incompatible with the MilliKelvin temperatures
needed for single spin measurement \cite{ladd05}. These arguments
suggest that \emph{any scheme for spectral diffusion control
  must rely on excitation of the qubits themselves, and not excitation
  of the nuclei}.

Here we consider the Carr-Purcell-Meiboom-Gill sequence (CPMG)
\cite{carr52,slichter96} as a scheme for spectral diffusion control.
The CPMG pulse sequence consists of the successive application of
qubit $\pi$-rotations perpendicular to $\bm{B}$ at time instants
$(2n+1)\tau$, $n=0,1,2,\ldots$. This procedure leads to the formation
of spin echoes at instants $(2n+2)\tau$, whose coherence is
substantially enhanced when compared to qubit evolution in the absence
of $\pi$-pulsing. Coherence enhancement takes place provided $\tau$
remains below a threshold $\tau_c$, the typical time scale for
environmental fluctuations.  As $\tau$ is decreased further, the
coherence gain can be substantially amplified, with an increased cost
in the number of $\pi$ pulses.  Here we develop a microscopic theory
which is able to predict, without any fitting parameters, the
inter-pulse time $\tau_{n}$ needed to provide $n$ times more coherence
for a spin qubit subject to nuclear-induced spectral diffusion in a
bath of general spin $I$. Our explicit calculations reveal that in
realistic devices nuclear spin noise can only be suppressed if $\tau$
is inversely proportional to the nuclear spin quantum number $I$
squared, establishing the overhead requirement for CPMG control of an
isolated spin coupled to a nuclear bath of general spin $I$.

Available spectral diffusion theories have focused solely on the case
of $I=1/2$ environments, using two-level telegraph noise models to
analyze the coherence time of a central spin-$1/2$
\cite{herzog56,desousa03b}. However many solid state spin-based
proposals rely on high spin environments, an important example being
the GaAs ($I=3/2$) quantum dot.  Here one often has to add In
($I=9/2$) to enable single spin read-out \cite{ono02,deng05} in
addition to a substantial amount of Al ($I=5/2$) to achieve
confinement \cite{kato03}.  InAs self assembled quantum dots are also
promising with regard to optical manipulation and transport
\cite{imamoglu03,ono02,chen04,ribeiro03}. The requirements for CPMG
control in these structures are not yet known. In this work we first
characterize spectral diffusion for $I>1/2$ environments and then
analyse coherence control with CPMG.  Our results establish CPMG as a
remedy for the absence of $I=0$ isotopes in III-V materials
\cite{tyryshkin03} as long as existing schemes for fast and precise
spin rotation are further developed \cite{kato03,chen04,note2}.

\section{Spectral diffusion decay induced by $I>1/2$ nuclear spins}

%Flip-flop rates
%Hamiltonian, pair of spin-I nuclei
Our qubit/spin $I$ bath Hamiltonian is given by
\begin{eqnarray}
{\cal H}&=&\gamma_{S}B S_{z}-\gamma_{I}B\sum_{n}I_{nz}\nonumber\\
&&+\sum_{n}A_{n}I_{nz}S_{z}
-4\sum_{n<m}b_{nm}I_{nz}I_{mz}\nonumber\\
&&+\sum_{n<m}b_{nm}(I_{n+}I_{m-}+I_{n-}I_{m+}).\label{htot}
\end{eqnarray}
The first term describes the qubit two-level structure, $\gamma_S$ and
$S_z$ being its gyromagnetic ratio and spin operator.  The nuclear
spin operators $\bm{I}_n$ are coupled to the qubit through the
parameters $A_n$. These are determined by hyperfine coupling (for an
electron spin qubit) or by dipolar interaction (for a nuclear spin
qubit).  The inter-nuclear dipolar coupling $b_{nm}$ is responsible
for nuclear spin fluctuations (for details see
Ref.~\onlinecite{desousa03b}).  In particular, the last term of
Eq.~(\ref{htot}) induces ``flip-flop'' transitions between pairs of
nuclear spins. When such a transition takes place, the qubit level
spacing shifts by $2 \Delta_{nm}=|A_n-A_m|$, leading to phase
randomization and to decoherence.  The dipolar coupled spin-$I$
lattice acts as a bath which is coupled to the qubit with strength
$\Delta_{nm}\equiv \Delta$.

In the semiclassical approach to spectral diffusion \cite{desousa03b},
the intricate nuclear spin evolution stemming from Eq.~(\ref{htot}) is
effectively described by a set of uncorrelated classical stochastic
transitions (flip-flops), with a characteristic rate $\Gamma$ obtained
from first principles. It is appropriate to use a sudden-jump model
for the description of these processes because the time scale for
thermal fluctuation is much shorter than $\frac{1}{\Gamma}$
($\hbar/k_BT\ll$~ns$\ll \frac{1}{\Gamma}\sim$~ms).  Correlation
between flip-flop events of fluctuating pairs located near each other
become important only when these pairs flip-flop several times within
$\tau$ (or equivalently $\Gamma \tau \gg 1$). For relevant time scales
($2\tau<T_2$) we can justify our neglect of flip-flop correlation {\it
  a posteriori} by noting that the condition $\Max{\Gamma} \tau \gg 1$
is never realized in the physical cases considered here.

\subsection{Calculation of the flip-flop transition rates in a
  spin-$I$ bath}

In the semiclassical theory, nuclear spin fluctuation is decoupled from
the qubit by setting $S_z\rightarrow 1/2$ in the full Hamiltonian
Eq.~(\ref{htot}). This leads to an effective Hamiltonian ${\cal
  H}^{'}$ for nuclear spin evolution under the inhomogeneous field
produced by the qubit,
\begin{equation} 
{\cal H}^{'}={\cal H}_{0}+\sum_{n<m} F_{nm}(t),
\end{equation}
\begin{equation} 
{\cal H}_{0}=-\sum_{n}\left(\gamma_{I}B - \frac{1}{2}A_{n}\right)
I_{nz} -4 \sum_{n<m}b_{nm}I_{nz}I_{mz},
\label{h0} 
\end{equation} 
\begin{equation} 
F_{nm}(t) = b_{nm} \left(I_{n+}I_{m-}\textrm{e}^{-i\omega
  t}+I_{n-}I_{m+}\textrm{e}^{i\omega t}\right).
\label{fnm}
\end{equation} 
In Eq.~(\ref{fnm}), we have introduced a fictitious frequency $\omega$
which allows a connection to the method of moments
\cite{desousa03b}. After we calculate the flip-flop rate, we will take
the limit $\omega\rightarrow 0$. 

Without loss of generality, we assume $n=1$, $m=2$.
The eigenstates of ${\cal H}_0$ are given by
$|M_1,M_2,\ldots\rangle$, with $M_{i}=-I,-I+1,\ldots,I$.
The flip-flop operators $F_{12}(t)$ do not connect
all states because they conserve ${\cal J}=(M_{1}+M_{2})/2$.  Hence
flip-flop dynamics break the pair Hilbert space into disconnected
subspaces labeled by ${\cal J}=I,I-1/2,\ldots,-I$. Each subspace has
dimension $2(I-|{\cal J}|)+1$ and can thus be alternatively labeled by
a pseudo-spin $I'=I-|{\cal J}|$, which is seen to correspond to
projection of the pseudo-spin operator $(\bm{I}_{1}-\bm{I}_{2})/2$
with projection index $M'=(M_1-M_2)/2=-I',\ldots,I'$.  Each nuclear
pair contains two subspaces for each $I'=0,1/2,1,\ldots,I-1/2$ (since
${\cal J}$ can be positive or negative) and one subspace with $I'=I$.

The Fermi Golden rule rate for a transition $M'\leftrightarrow M'-1$ 
between the state $|M'\rangle = |M_1,M_2,M_3,\ldots,M_N\rangle$ and 
$|M'-1\rangle = |M_1-1,M_2+1,M_3,\ldots,M_N\rangle$
is given by 
\begin{eqnarray}
\Gamma &=& 2\pi b_{12}^2\sum_{M_3,\ldots,M_N}
p(M_3,\ldots,M_N) \nonumber\\
&&\!\!\!\!\!\times\left|
\langle M'-1 | I_{1-}I_{2+} | M'\rangle\right|^2 \delta(E_{M'-1}^{0}-E_{M'}^{0}-\omega),
\label{fgr}
\end{eqnarray}
where $p(M_3,\ldots)$ are Boltzmann probabilities for the unperturbed
energies $E_{M'}^{0}$. It is convenient to define an auxiliary
function
\begin{equation}
\rho(\omega) = \sum_{M_3,\ldots,M_N}
p(M_3,\ldots,M_N)\delta(E_{M'-1}^{0}-E_{M'}^{0}-\omega),
\label{rho}
\end{equation}
which can be interpreted as a density of
states. After evaluating the matrix element in Eq.~(\ref{fgr}) we get 
\begin{eqnarray}
\Gamma&=&2\pi b_{12}^2 (2I-I'+M')(2I-I'-M'+1)\nonumber\\
&&\times
(I'+M')
(I'-M'+1)\rho(\omega),
\end{eqnarray}
where we recall that $I'=I-|M_1+M_2|/2$ and $M'=(M_1-M_2)/2$. We can
now calculate explicitly the moments of the function
$\rho(\omega)$. For example, $\int \rho(\omega)d\omega = 1$, while the
first moment is given by
\begin{eqnarray}
\overline{\omega}_{12}&=&\int_{-\infty}^{\infty} \omega\rho(\omega)d\omega \nonumber\\
&=& \sum_{M_3,\ldots,M_N}
p(M_3,\ldots,M_N)(E_{M'-1}^0-E_{M'}^0)\nonumber\\
&=& \frac{1}{2}(A_2-A_1)-4b_{12}(M_1-M_2-1)\nonumber\\
&&+4\sum_{k\neq 1,2}
(b_{1k}-b_{2k})\langle m_k\rangle.
\end{eqnarray}
Here the thermal average $\langle m_k\rangle = \partial
\ln{Z}/\partial \beta'$ can be calculated from the partition function 
\begin{equation}
Z=\sum_{m=-I,\ldots,I} \textrm{e}^{\beta' m},
\end{equation}
with $\beta'=\hbar \gamma_I B / k_B T$ (for $B\gg I
\sum_n b_{nm}/\gamma_I\sim 10$~G we can neglect the dipolar energy in the
Boltzmann distributions). 
The second moment is given by 
\begin{eqnarray}
\kappa_{12}^2&=&\int(\omega -\overline{\omega}_{12})^2 \rho(\omega)d\omega \nonumber\\
&=& 16 \sum_{k\neq 1,2} (b_{1k}-b_{2k})^2 \langle (m_k-\langle m_k\rangle)^2\rangle,
\end{eqnarray}
where $\langle(m_k-\langle m_k\rangle)^2\rangle=\partial^2
\ln{Z}/\partial^2\beta'$. The fourth moment $q_{12}^4$ 
can also be calculated,
allowing us to show that the ratio $q_{12}^4/(3\kappa_{12}^2)\sim
1$. Thus the Gaussian function provides a reasonable fit to Eq.~(\ref{rho}),
\begin{equation}
\rho(\omega)\approx \frac{1}{\sqrt{2\pi \kappa_{12}^2}}
\exp{\left[-\frac{\left(\omega-\overline{\omega}_{12}\right)^2}{2\kappa_{12}^2}\right]}.
\end{equation}
The final expression for the flip-flop rate for a nuclear spin
pair $(n,m)$ is \cite{note0}
\begin{eqnarray}
\Gamma_{M'} &=& (2I-I'+M')(2I-I'-M'+1)(I'+M')
\nonumber\\
&&\times (I'-M'+1)
\frac{\sqrt{2\pi} b_{nm}^2}{\kappa_{nm}}
\exp{\left(-\frac{\omega_{nm}^2}{2\kappa_{nm}^2}\right)}.
\label{tnm}
\end{eqnarray}
When $\beta'\ll 1$, there is only a weak temperature and $B$ field
dependence in Eq.~(\ref{tnm}). For example, $T=100$~mK and $B=10$~T
leads to $\beta' \sim 0.1$. In this approximation we have 
$\langle m_k\rangle\approx 0$ and $\langle(m_k-\langle
m_k\rangle)^2\approx I(I+1)/3$. This leads to the approximate expressions
\begin{equation}
\overline{\omega}_{nm}\approx\Delta_{nm} - 4b_{nm}(2M'-1),
\label{omn}
\end{equation}
\begin{equation}
\kappa_{nm}^{2}\approx\frac{16}{3}I(I+1)\sum_{i\neq n,m}
\left(b_{ni}-b_{mi}\right)^2.\label{kappa}
\end{equation}
$\kappa_{nm}$ represents the amount of energy the dipolar system can
supply for a nuclear flip-flop to take place.  Nuclear spins located
near the center of the electron wave function are characterized by
$\overline{\omega}_{nm}\gg \kappa_{nm}$, with exponentially small
flip-flop rates.  These are said to form a frozen core which does not
contribute to spectral diffusion (a similar effect has been reported
in optical experiments, see Ref.~\onlinecite{devoe81}).  On the other
hand nuclei satisfying $\overline{\omega}_{nm}\sim \kappa_{nm}$
possess appreciable flip-flop rates together with a sizable hyperfine
shift $\Delta_{nm}$.  These are located in a shell around the center of the
electronic wave function, and are responsible for most of the spectral
diffusion coherence decay \cite{desousa03a,desousa03b}.  Below we show
that to a very good approximation the $M'$ dependence in
Eq.~(\ref{tnm}) averages out, so that we may consider just the average
rate
\begin{equation}
\Gamma(I',I)=\langle\Gamma_{M'}\rangle
\approx\xi(I',I)\;\Gamma(I,I)\sim II'^{2},
\label{gii}
\end{equation}
with the parameter $\xi(I',I)$ connecting different subspaces of a
spin-$I$ pair,
\begin{equation}
\xi(I',I)=\frac{I'(I'+1)\left[1+5I(2I+1)+I'(2I'-10I-3)\right]}{I(I+1)
\left[1+2I(I+1)\right]}.\label{xi}
\end{equation}

\begin{figure}
\includegraphics[width=3.4in]{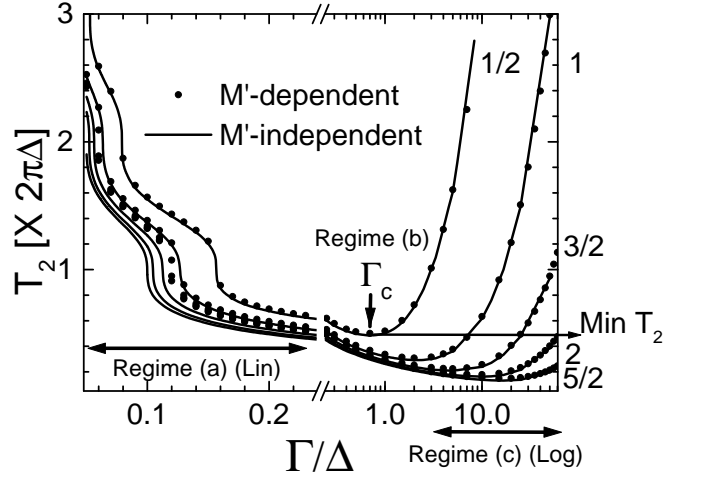}
\caption{
  Behavior of $1/e$ decay times for the first CPMG echo in pseudo-spin
  subspaces $I'\leq 5/2$ as a function of flip-flop rate $\Gamma$.
  Circles: calculation with $M'$-dependent rates $\Gamma_{M'}$.  Solid
  lines: calculation with average rate $\langle \Gamma_{M'} \rangle$.
  Each curve achieves minimum coherence at $\Gamma=\Gamma_{\rm{c}}$,
  which is identified as a critical correlation rate. Small $\Gamma$
  (or strong coupling $\Delta$) imply low frequency noise with time
  correlation [regime(a)], while large $\Gamma$ (or weak coupling
  $\Delta$) imply motional narrowing or white noise behavior [regime
  (c)].  Note the strong $I'$ dependence in the weak coupling region.
  Here we have $\Min{(T_{2})}\approx 6/\Delta(2I'+1)$, and
  $\Gamma_{\rm{c}}\approx (2I'+1)^3/12$.\label{fig1}}
\end{figure}

\subsection{Random walk theory for the fluctuation of a $I'$ subspace}

% Random walk theory
%Formulation/solution of Stochastic model
We now formulate a random walk theory for the coupling of the
qubit to the flip-flop dynamics in each pseudo-spin subspace $I'$ and
then construct the coherence decay as a weighted average over all
pseudo-spin subspaces.  Our model assumes that each $I'$ subspace
fluctuates independently, i.e. pairs are uncorrelated. As a result,
the qubit Zeeman frequency is subjected to several random fields of
the form $\Delta\sigma_{I'}(t)$, where $\sigma_{I'}$ is a hard wall
random walk variable assuming the values
\begin{equation}
\sigma_{I'}(t)= \left\{-2I',-2(I'-1),\ldots, 2I'\right\}.\label{sigma}
\end{equation}
The theory for qubit decoherence resulting from the two $I'=1/2$
subspaces is described in Ref.~\onlinecite{desousa03b}. It has a simple
structure because for $I'=1/2$, $\sigma_{I'}(t)$ can be written as
$(-1)^{N(t)}$, with $N(t)$ a Poisson random variable with parameter
$\Gamma t$. The generalization of this telegraph process to the
multi-level noise present when $I' > 1/2$ is achieved by defining an
occupation probability vector $\bm{p}(t)$ and its Markovian evolution
operator $U(t)$:
\begin{eqnarray}
\bm{p}(t)& = & (p_{-I'},p_{-I'+1},\ldots,p_{I'}), \\
         & = & \exp{\left(-t\bm{A}\right)}\cdot \bm{p}(0) \\
         & \equiv & \bm{U}(t)\cdot\bm{p}(0).
\label{pt}
\end{eqnarray}
For first-order flip-flop transitions we have
\begin{equation}
\dot{p}_i=\Gamma_i\; p_{i-1}-(\Gamma_i+\Gamma_{i+1})p_{i}
+\Gamma_{i+1}\;p_{i+1},
\label{pdot}
\end{equation}
\begin{equation}
A_{ij}=-\Gamma_i\delta_{i-1,j}+(\Gamma_i+\Gamma_{i+1})\delta_{i,j}-
\Gamma_{i+1}\delta_{i+1,j}.
\label{ma}
\end{equation}
We can determine the echo decay for all Markovian processes described by
$\bm{A}$ using the time-ordered correlation function ($t_k\geq
t_{k-1}\geq\cdots\geq t_1$)
\begin{eqnarray}
&&\langle\sigma_{I'}(t_k)\cdots\sigma_{I'}(t_1)\rangle= 
(1\cdots 1)\cdot \bm{\Sigma}_{I'}
\cdot \bm{U}(t_{k}-t_{k-1})\nonumber\\
&&\cdot\bm{\Sigma}_{I'}\cdot
\bm{U}(t_{k-1}-t_{k-2})\cdots \bm{\Sigma_{I'}}\cdot \bm{U}(t_1)\cdot \bm{p}(0),
\label{tcor}
\end{eqnarray}
defined by correlations between values of the spin state matrix 
\begin{equation}
\bm{\Sigma}_{I'}=\diag{\left\{-2I',-2(I'-1),\ldots,2I'
\right\}}
\label{zi}
\end{equation}
at times $t_1, t_2, \ldots, t_k$.
The Hahn echo is then calculated using a double average
\begin{equation}
v_{I'}(2\tau)=\left\langle\left\langle \exp{\left( i
        \int_{0}^{2\tau}s(t)\sigma_{I'}(t)dt
\right)}  \right\rangle_{\sigma_{I'}(0)=\sigma} \right\rangle_{\sigma},
\label{echo_avg}
\end{equation}
where the echo function is $s(t)=1$ for $t\leq \tau$ and $s(t)=-1$ for
$t>\tau$ (we set $\Delta=1$ for simplicity, since it will be recovered
later as the unit of time). The inner average is over $\sigma_{I'}(t)$
trajectories starting from $\sigma_{I'}(0)=\sigma$.  The outer average
is a thermal average over all possible initial states $\sigma$.  The
n-th echo of the CPMG sequence can then be simply obtained from the
n-th power of Eq.~(\ref{echo_avg}) \cite{slichter96,note3}.

The difficulty of evaluating Eq.~(\ref{echo_avg}) lies in ensuring a
proper treatment of $s(t)$.  We propose here a matrix method that
enables the incorporation of any step-wise constant $s(t)$ explicitly.
After expanding and rearranging the integrals we obtain
\begin{eqnarray}
v_{I'}(2\tau)&=&\sum_{\sigma}p_{\sigma}(0)
\sum_{j,k=0}^{\infty}(-i)^{j}(i)^k \int_{\tau}^{2\tau}dt_{j}'\ldots
\int_{\tau}^{t_2'}dt_1' \nonumber\\
&&\int_{0}^{\tau}dt_{k}\ldots\int_{0}^{t_2}dt_1 
\left\langle\left[ \sigma(t_j')\cdots \sigma(t_1')\right]\right.\nonumber\\
&&\left.\times\left[\sigma(t_k)\cdots\sigma(t_1)\right]\right\rangle_{\sigma},
\label{vsum}
\end{eqnarray}
where $\tau\geq t_k\geq\ldots\geq t_1\geq 0$ and $2\tau\geq
t_j'\geq\ldots\geq t_1'\geq \tau$ are partitions before and after the
$\pi$ pulse. Using Eq.~(\ref{tcor}) and the Markovian identity 
$\bm{U}(t_1'-t_k)=\bm{U}(t_1'-\tau)\cdot\bm{U}(\tau-t_k)$, Eq.~(\ref{vsum}) can
be written as
\begin{equation}
v_{I'}(2\tau)=\frac{1}{2I'+1}(1,\ldots,1)\cdot\bm{M}_{-}\cdot\bm{M}_{+}\cdot
\left(
\begin{array}{c}
1\\
\vdots\\
1
\end{array}
\right),\label{vmatrix}
\end{equation}
where the initial probabilities are simply $p_{\sigma}(0)=1/(2I'+1)$.
The matrices $\bm{M}_{\pm}$ are given by (after the substitution
$t_{j}''=t_j'-\tau$)
\begin{equation}
\bm{M}_{-}(\tau)=\sum_{j=0}^{\infty} (-i)^j \int_{0}^{\tau}dt_j'' \bm{m}_{j}(t_j''),
\label{mm}
\end{equation}
\begin{equation}
\bm{M}_{+}(\tau)=\sum_{k=0}^{\infty} (i)^k \int_{0}^{\tau}dt_k \bm{U}(\tau-t_k)\cdot \bm{m}_{k}(t_k),
\label{mp}
\end{equation}
where $\bm{m}_{l}(t_l)$ is obtained from the recurrence relation
\begin{equation}
\bm{m}_{l}(t_l)=\int_{0}^{t_l}dt_{l-1} \bm{\Sigma}_{I'}\cdot
\bm{U}(t_l-t_{l-1})\cdot\bm{m}_{l-1}(t_{l-1}),
\end{equation}
\begin{equation}
\bm{m}_{1}(t_1)=\bm{\Sigma}_{I'}\cdot\bm{U}(t_1).
\end{equation}
Going to Laplace space the convolution integrals are simply converted 
to multiplications, 
\begin{equation}
\bm{\widetilde{m}}_{l}(r)=\int_{0}^{\infty}\textrm{e}^{-rt}\bm{m}_l(t)dt=\left[\bm{\Sigma}_{I'}
\cdot\bm{\widetilde{U}}(r)\right]^l.
\end{equation}
Using Eqs.~(\ref{mm}), (\ref{mp}) we obtain closed expressions
for the Laplace transforms of the matrices $\bm{M}_{\pm}$,
\begin{equation}
\bm{\widetilde{M}_{-}}(r)=\frac{1}{r}\frac{\bm{1}}
{\bm{1}+i\bm{\Sigma}_{I'}\cdot\bm{\widetilde{U}}(r)},
\label{mmlap}
\end{equation}
\begin{equation}
\bm{\widetilde{M}_{+}}(r)=\bm{\widetilde{U}}(r)\frac{\bm{1}}
{\bm{1}-i\bm{\Sigma}_{I'}\cdot\bm{\widetilde{U}}(r)}.
\label{mplap}
\end{equation}
These general expressions for multi-level Markovian echo can be
further simplified when written as a function of the eigenvectors and
eigenvalues of $\bm{A}$. 

%Numerical results for I' / figure containing T_2 vs Gamma.
Solving Eqs.~(\ref{vmatrix}), (\ref{mmlap})-(\ref{mplap}), using
(\ref{pt})-(\ref{ma}), allows evaluation of the decoherence time $T_2$
from the $1/e$ decay of the Hahn echo, Eq.~(\ref{vmatrix}).  An
analytical expression for the Hahn echo is obtained using {\it
  Mathematica} \cite{mathematica51}.  Fig.~\ref{fig1} shows $T_2$ as a
function of average flip-flop rate $\Gamma$ for several $I'$
subspaces.  First we note that for all $I'$, averaging over the $M'$
dependences in $\Gamma_{M'}$ is an excellent approximation [resulting
from the outer average in Eq.~(\ref{echo_avg})].  For each $I'$, we
identify a critical bath correlation rate $\Gamma_{\rm{c}}$ for which
the minimum decay time $T_2$ is found.  This leads to a natural
division into three different regimes which can be seen to possess
different capabilities for coherence enhancement by CPMG sequences.
We recall that application of a CPMG sequence to Hahn echo given by
$\exp{[-(2\tau/T_2)^d]}$ transforms this to
$\exp{(-2n\tau/T_{\rm{CPMG}})}$ where
$T_{\rm{CPMG}}=[T_2/(2\tau)]^{d-1}T_2$ provides a measurement of the
coherence enhancement which is obtained for all $d>1$ (no enhancement
results for $d=1$)\cite{slichter96}. 

We refer to Fig.~\ref{fig1} for an illustration of regimes (a), (b),
(c).  Regime (a) holds for $\Gamma/\Delta\ll \Gamma_{\rm{c}}$.  This
is the regime of sudden-jump spectral diffusion (strong coupling
$\Delta$) which has oscillations in the echo envelope due to
precession under a finite $\Delta$. Assuming that at most two
flip-flops take place, we obtain an expansion up to second order in
$(\Gamma/\Delta)$,
\begin{eqnarray}
v_{I'}(2\tau)&&\approx
\frac{2\textrm{e}^{-\rho}}{2I'+1}
\left\{
1+\frac{2I'-1}{2}\textrm{e}^{-\rho}+\left(\frac{\Gamma}{\Delta}\right)
\sin{\delta}
\right.\nonumber\\
&&\times\left.\left[1
+(2I'-1)\textrm{e}^{-\rho}\right]
+\left(\frac{\Gamma}{\Delta}\right)^2 \left[1-\cos{\delta}
\right.\right.\nonumber\\
&&\left.\left.
+\frac{2I'-1}{2}\textrm{e}^{-\rho}\left(1-\cos{2\delta}\right)\right]
\biggr\}\right.,
\label{vipslow}
\end{eqnarray}  
where $\rho=2\tau\Gamma$ and $\delta=2\tau\Delta$.  When
$\tau<1/\Delta$, CPMG provides coherence enhancement proportional to
$(T_2/\tau)^2$.  Regime (b) holds for $\Gamma/\Delta\sim
\Gamma_{\rm{c}}$.  This is the regime of critical spectral diffusion,
which occurs when $T_2$ reaches a minimum.  The minimum is a result of
the inevitable mixing of the random walk noise in a finite walk, such
as that in the pseudo-spin subspace considered here.  Coherence
enhancement is approximately proportional to $(T_2/\tau)$ for
$\tau<1/\Delta$.  Finally, regime (c) satisfies $\Gamma/\Delta\gg
\Gamma_{\rm{c}}$.  This corresponds to the continuum or weak coupling
regime. For $2\tau\sim T_2$ nuclear spin fluctuation is so fast that
the coherence of the central spin-$1/2$ is enhanced.  Motional
narrowing takes place when the stochastic process traverses the random
walk space several times; hence it is insensitive to all eigenvalues
of $\bm{A}$ except for the lowest non-zero eigenvalue $a^*$ and
associated eigenvector $\bm{v^*}$.  The latter observation allows us
to derive an analytical expression for the decoherence time at
motional narrowing, when qubit spin echo $v_{I'}\approx
\exp{(-2\tau/T_{2})}$ with
\begin{equation}
T_{2}\approx 
\frac{2I'+1}{2I'}\frac{a^*}{v_{1}^*\sum v_{j}^* (\bm{\Sigma}_{I'})_{jj}}
\frac{\Gamma}{\Delta^{2}}
\propto I'^{-4}.
\label{tmmn}
\end{equation}
When Eq.~(\ref{tmmn}) holds, no CPMG enhancement is observed.  We note
however that as $\tau$ decreases below a certain threshold value
$\tau_c\approx 1/(a^*\Gamma) \approx (2I'+1)^2/(8\Gamma)$, crossover
to $v_{I'}\sim \exp{(-\tau^3)}$ behavior takes place and
consequently the CPMG pulse sequence will again provide effective
enhancement of coherence. The threshold value $\tau_c$ is related to
continuous Gaussian noise, since the limit $\Gamma\rightarrow\infty$,
$\Delta\rightarrow 0$ and $I'\rightarrow\infty$ with $\tau_c$ and
$I'\Delta$ constant is given by Eq.~(A11) of Ref.~\onlinecite{klauder62}.
\begin{figure}
\includegraphics[width=3.4in]{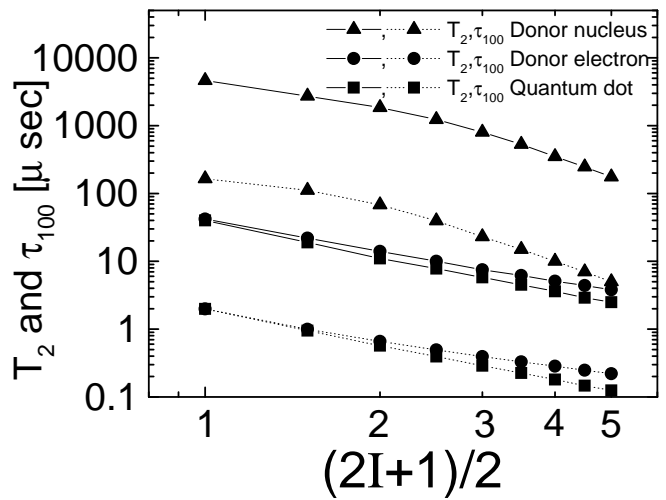}
\caption{Coherence times $T_2$ and inter-pulse times $\tau_{100}$
  required for 100$\times$ coherence enhancement as a function of
  nuclear spin quantum number $I$. We find that both $T_2$ and
  $\tau_{100}$ scale as $\sim 1/I^{1.5}$.}
\label{fig2}
\end{figure}

In order to determine the dependence of spin coherence on the bath
nuclear spin value $I$, we make an average of Eq.~(\ref{vmatrix}) over
all $I'$ subspaces and nuclear species s. We recall that most
materials considered here are a mixture of different isotopes. In a
strong external field, only flip-flop between the same nuclear species
needs to be considered, since flip-flop between different species is
strongly suppressed due to an offset in hyperfine shifts.  The
resulting echo envelope is
\begin{equation}
\langle v\rangle_{I}(2\tau) = \sum_{s,I'}p(s,I')
v_{I'}[2\tau, \xi(I',I)\Gamma],
\label{v_avg}
\end{equation}
where $p(s,I')\propto f_{s}^{2}(2I'+1)\cosh{[2(I-I')\hbar\gamma_I
  B/k_B T]}$ with $f_s$ the isotopic abundance for nuclear species s.
Eq.~(\ref{v_avg}) shows that substantial nuclear spin polarization (or
extremely low nuclear spin temperature $T$) would be required to
suppress spectral diffusion. For example, $p(0)=0.9$ (corresponding to
90\% of nuclear spins having maximum magnetization $I$) increases
$T_2$ by a factor of $2$ only, while for the completely unpolarized
case, $p(I')=2(2I'+1)/(2I+1)^2$. \cite{dassarma05}  Numerical
evaluation shows that the dependence of Eq.~(\ref{v_avg}) on
$\Gamma/\Delta$ is qualitatively similar to that seen for the
individual $I'$ subspaces in Fig.~\ref{fig1}, but with different
scaling behavior for the characteristic quantities. For example,
$\Gamma_{\rm{c}}\approx [(2I+1)/3]^{2.7}$, allowing possible crossover
between the three regimes as $I$ increases.

\section{Numerical results}

The complete coherence decay is obtained as a product of decays from
multiple fluctuator pairs $n,m$.  Most $I > 1/2$ materials of interest
have similar dipolar couplings ($b_{nm}$) so that we may conclude that
CPMG control depends primarily on the shape of the qubit wave function
(which determines the distribution of $\Delta_{nm}$) and on the
magnitude of the nuclear spin $I$. Fig.~\ref{fig2} shows our numerical
results for $T_2$ and $\tau_{100}$ (inter-pulse time needed for
$100\times$ coherence enhancement) for the case of donor impurities
(electron and nuclear spin qubit) and a III-V quantum dot (electron
spin qubit -- see Ref.~\onlinecite{elzerman04}).  For an electron spin it turns
out most nuclear pairs are fluctuating in regime (a) with $\Gamma
T_2\ll 1$, implying the threshold for coherence enhancement $\tau_c$
is quite close to $T_2$. As $\tau$ decreases below $\tau_c\sim T_2$,
coherence enhancement scales as $(T_2/\tau)^2$, yielding
$\tau_{100}\sim 0.05 T_2$.  On the other hand, coherence decay of
nuclear spin qubits is sensitive to fluctuators in regimes (a)--(c)
with $\Gamma T_2\lesssim 1$, which makes the threshold $\tau_c$
somewhat smaller than $T_2$ \cite{note1}. Here $\tau_{100}\sim 0.03
T_2$, showing a smooth crossover from $1/I$ to $1/I^3$.  Our study
shows that the application of $\pi$-rotations every $\tau \sim
0.1$~$\mu$s for an electron spin qubit and $\tau\sim 10$~$\mu$s for a
nuclear spin qubit is sufficient to increase coherence by two orders
of magnitude, effectively reaching the phonon emission limit
\cite{elzerman04}.

\section{Discussion and conclusion}

Fig.~\ref{fig2} answers the previously open question of the dependence
of spectral diffusion with increasing $I$. There had been earlier
speculation \cite{desousa03thesis} that the spectral diffusion rate
$1/T_2$ might decrease with increasing $I$ because of motional
narrowing [due to faster flip-flops - notice that averaging over the
$I'$ subspaces in Eq.~(\ref{gii})results in $\Gamma\propto I^3$]. We
show here (see Fig.~\ref{fig1}) that the drastic increase in size of
the Hilbert space for increasing $I$ does actually compensate for the
strong amplification of flip-flop rates and that therefore the system
does not enter the motional narrowing regime, resulting instead in
more spectral diffusion induced decoherence as $I$ increases.

Another interesting point to note is that the repetition rate needed
to suppress spectral diffusion (of the order of $I^2\times$MHz) is
significantly different than that corresponding to the strength of
dipole-dipole interactions, which are given by the root mean square of
the last two terms of Eq.~\ref{htot}. This root mean square average is
commonly referred to as the square root of the second moment (see,
e.g., Eq.~3.53 of Ref.~\onlinecite{slichter96}).  For the systems of
interest this is of the order of $I\times$KHz. The rate for
controlling spectral diffusion is \emph{significantly faster} than
dipole-dipole coupling because hyperfine coupling (absent from the
second moment equation) is very strong (up to MHz) and the spectral
diffusion decay occurs due to the collective fluctuation of several
nuclei under the field of the electron.  We emphasize that the
relative time scale for CPMG control of spectral diffusion and its
dependence on the nuclear spin value $I$ that we have established here
can only be derived by carefully considering the combined effect of
hyperfine interaction and collective dipolar fluctuation of $~10^4$
nuclear spins under the inhomogeneous hyperfine field produced by the
electron, as we have done in this paper.

The effectiveness of the CPMG sequence in controlling spectral
diffusion depends largely on the availability of fast methods for
precise spin manipulation.  The pulsing time has to be much smaller
than $\tau$, while the spin rotation angle should be precisely tuned
to $\pi$ \cite{morton05}.  The train of $\pi$ pulses must also be
applied without heating the sample above the required $\sim 100$~mK
temperature. Currently, there are several proposals for overcoming
this technical challenge, including all electrical spin resonance
\cite{kato03}, optical manipulation \cite{chen04}, and implanting a
low power microwave source nearby the qubit \cite{friesen04}.  If this
technical problem can be solved, our work indicates that the CPMG
sequence will play a crucial role in extending electron spin coherence
times in semiconductors.

In conclusion, we have provided a scheme for coherence control of
localized spins in semiconductors subject to general nuclear spin-$I$
fluctuations. We find that the rate for $\pi$-pulsing needed for
substantial coherence enhancement is $\sim I^2$, yielding an efficient
route for achieving long time spin coherence in semiconductors using
advanced pulse technology.

We acknowledge support from the DARPA SPINS
program and ONR under Grant No. FDN0014-01-1-0826.

%
%%%%%%%%%%%%%%%%%%%%%%%%%%%%%%%%%%%%%%%%%%%%%%%%%%%%%%%%%%%%%%%
%


\begin{thebibliography}{99}
\bibitem{note0} Eq.~(\ref{tnm}) supersedes Eq.~(46) in
Ref.~\onlinecite{desousa03b}, where the independence of different $I'$
subspaces was not taken into account. Also, the correct averaging is
given by Eq.~(\ref{v_avg}) instead of Eq.~(59) in
Ref.~\onlinecite{desousa03b}. After including all these corrections we
noted that the numerical results for $T_2$ in
Ref.~\onlinecite{desousa03b} increased by about 10\%.
\bibitem{note3} For a single (Hahn) echo, Eq.~(\ref{echo_avg}) results
in the explicit expression Eq.~(\ref{vmatrix}). It is straightforward
to generalize Eq.~(\ref{vmatrix}) for a train of $\pi$-pulses,
allowing exact evaluation of the n-th CPMG echo. We verified that this
exact result is very well approximated by the n-th power of
Eq.~(\ref{vmatrix}), as noted in Eq.~8.4 of
Ref.~\onlinecite{slichter96}.
\bibitem{note1} Nuclear and electron spin of a donor impurity are
subject to the same $\Gamma_{nm}$, but $\Delta_{nm}$ is $\sim 10^3$
smaller for the former.
\end{thebibliography}
\end{document}